\documentclass[showpacs,preprintnumbers,amsmath,amssymb,prd]{revtex4-1}

\usepackage{graphicx}
\usepackage{epsfig}

\def\be{\begin{equation}}
\def\ee{\end{equation}}
\def\bea{\begin{eqnarray}}
\def\eea{\end{eqnarray}}
\def\ba{\begin{array}}
\def\ea{\end{array}}

\begin{document}
\title{Quark sea asymmetries of the octet baryons}
\author{Neetika Sharma and Harleen Dahiya}
\affiliation{Department of Physics,\\ Dr. B.R. Ambedkar National
Institute of Technology,\\ Jalandhar, 144011, India}

\begin{abstract}

The effects of  ``quark sea'' in determining the flavor structure
of the octet baryons have been investigated in the chiral
constituent quark model ($\chi$CQM). The $\chi$CQM is able to
qualitatively generate the requisite amount of quark sea and is
also known to provide a satisfactory explanation of the proton
spin and related issues in the nonperturbative regime. The Bjorken
scaling variable $x$ has been included phenomenologically in the
sea quark distribution functions to understand its implications on
the quark sea asymmetries like $\bar d(x)-\bar u(x)$, $\bar
d(x)/\bar u(x)$ and Gottfried integral for the octet baryons. The
results strengthen the importance of quark sea at lower values of
$x$.
\end{abstract}

\maketitle

After the first direct evidence for the point-like constituents in
the nucleon \cite{point-like}, identified as the valence quarks
with spin-1/2 in the naive constituent quark model (NQM)
\cite{dgg,Isgur,yaouanc}, a lot of experiments have been conducted
to probe the structure of the proton in the deep inelastic
scattering (DIS) experiments. Surprisingly, the DIS results in the
early 80's \cite{emc} indicated that the valence quarks of the
proton carry only about 30\% of its spin and is referred to as the
``proton spin crisis'' in the NQM. These results provided the
first evidence for the proton being composed of three valence
quarks surrounded by an indistinct sea of quark-antiquark pairs
(henceforth referred to as the ``quark sea'').  In the present
day, the study of the composition of hadrons can be said to be
primarily the study of the quark sea and gluons and is considered
as one of the active areas in hadronic physics.

The conventional expectation that the quark sea perhaps can be
obtained through the perturbative production of the
quark-antiquark pairs by gluons produces nearly equal numbers of
$\bar u$ and $\bar d$. Until early 90's a symmetric sea w.r.t.
$\bar u$ and $\bar d$ was assumed, however, the famous New Muon
Collaboration in 1991 \cite{nmc} established the quark sea
asymmetry of the unpolarized quarks in the case of nucleon by
measuring $\bar d-\bar u$ giving first clear evidence for the
nonperturbative origin of the quark sea. This was later confirmed
by the Drell-Yan experiments \cite{baldit} which measured  a large
quark sea asymmetry ratio $\bar d/\bar u$ reminding us that the
study of the quark sea is intrinsically a nonperturbative
phenomena and it is still a big challenge to perform these
calculations from the first principles of QCD.

One approach to account for the observed quark sea asymmetry is
the pion cloud mechanism \cite{pioncloud} where the quark sea is
believed to originate from process such as virtual pion
production. It is suggested that in the deep inelastic
lepton-nucleon scattering, the lepton probe also scatters off the
pion cloud surrounding the target proton. The $\pi^+ (\bar d u)$
cloud, dominant in the process $p \rightarrow \pi^+ n$, leads to
an excess of $\bar d$ sea. However, this effect should be
significantly reduced by the emissions such as $p \rightarrow
{\Delta}^{++} + \pi^-$ with  $\pi^- (\bar u d)$ cloud. Therefore,
the pion cloud idea is not able to explain the significant $\bar d
> \bar u$ asymmetry. This approach can be improved upon by
adopting a mechanism which operates in the {\it interior} of the
hadron.

The chiral constituent quark model ($\chi$CQM) \cite{manohar} can
yield an adequate description of the quark sea generation through
the chiral fluctuations. The basic idea is based on the
possibility that chiral symmetry breaking takes place at a
distance scale much smaller than the confinement scale. In this
region, the effective degrees of freedom are the valence quarks
and the internal Goldstone bosons (GBs) which are coupled to the
valence quarks \cite{eichten,cheng,song} allowing a simple and
intuitive method to investigate the principle features of the
hadron structure. In the case of spin dependent quantities, the
$\chi$CQM is not only successful in giving a satisfactory
explanation of ``proton spin crisis'' \cite{eichten,hd} but is
also able to account for the baryon magnetic moments
\cite{hdorbit} and hyperon $\beta-$decay parameters
\cite{cheng,johan,ns}. However, in the case of quark distribution
functions, the latest developments by the NuSea (E866) \cite{e866}
and HERMES \cite{hermes} to determine the variation of the
sea-antiquark ratio $\bar d(x)/\bar u(x)$ and the difference $\bar
d(x)-\bar u(x)$ with Bjorken scaling variable $x$ have renewed
considerable interest in the quark sea asymmetries.

Recently, there has been substantial theoretical  progress to take
into account the effects of quark sea in determining the flavor
structure of the baryons and the question of sea asymmetry has
been investigated  by several authors using various
phenomenological models. Calculations have been carried out in the
meson cloud models \cite{mesoncloud}, chiral quark-soliton model
\cite{cqsm}, effective chiral quark model \cite{eccm}, statistical
models \cite{stat}, bag model \cite{bag}, model for parton
densities \cite{alwall}, radiative parton model \cite{reya} etc..
However, the inclusion of $x-$dependence has not yet been
successfully included in the quark distribution functions.
Therefore, pending further experiments, it would be interesting to
examine the flavor structure of the octet baryons at low energy,
thereby giving vital clues to the nonperturbative effects of QCD.
The study of $x-$dependence in the quark distribution functions
becomes particularly interesting for the $\chi$CQM where the
effects of quark sea and valence quarks can separately be
calculated.

The purpose of the present communication is to determine the sea
quark distribution functions and their asymmetries in the octet
baryons by phenomenologically incorporating $x-$dependence in the
$\chi$CQM. The extent of contributions coming from the different
sea quarks for the octet baryons can also be compared. To
understand the relation of the Bjorken scaling variable and quark
sea, it would be significant to study its implications in the
region $x<0.3$ which is a relatively clean region to test the
quark sea structure as well as to estimate their structure
functions and related quantities \cite{cdhs}.

The key to understand the ``proton spin crisis'', in the $\chi$CQM
formalism \cite{cheng}, is the fluctuation process \be q^{\pm}
\rightarrow {\rm GB} + q^{' \mp} \rightarrow  (q \bar q^{'})
+q^{'\mp}\,,
\label{basic} \ee $q \bar q^{'} +q^{'}$ constitute the ``quark sea''
\cite{cheng,song,hd,johan}. The effective Lagrangian describing
interaction between quarks and a nonet of GBs,  can be expressed
as \be {\cal L}= g_8 {\bf \bar
q}\left(\Phi+\zeta\frac{\eta'}{\sqrt 3}I \right) {\bf q}=g_8 {\bf
\bar q}\left(\Phi' \right) {\bf q}\,,\label{lag} \ee where
$\zeta=g_1/g_8$, $g_1$ and $g_8$ are the coupling constants for
the singlet and octet GBs, respectively, $I$ is the $3\times 3$
identity matrix. In terms of the SU(3) and axial U(1) symmetry
breaking parameters,  introduced by considering $M_s > M_{u,d}$,
$M_{K,\eta}> M_{\pi}$ and $M_{\eta^{'}} > M_{K,\eta}$
\cite{cheng,song,johan}, the GB field can be expressed as \bea
\Phi' = \left( \ba{ccc} \frac{\pi^0}{\sqrt 2} +
\beta\frac{\eta}{\sqrt 6} + \zeta\frac{\eta^{'}}{\sqrt 3} & \pi^+&
\alpha K^+ \\ \pi^- & -\frac{\pi^0}{\sqrt 2} +\beta
\frac{\eta}{\sqrt 6} +\zeta\frac{\eta^{'}}{\sqrt 3}  &  \alpha K^0
\\\alpha K^-  & \alpha \bar{K}^0  &  -\beta \frac{2\eta}{\sqrt 6}
+\zeta\frac{\eta^{'}}{\sqrt 3} \ea \right) {\rm and} ~~~~ q
=\left( \ba{c} u \\ d \\ s \ea \right)\,. \eea The parameter
$a(=|g_8|^2$) denotes the probability of chiral fluctuation  $u(d)
\rightarrow d(u) + \pi^{+(-)}$, whereas $\alpha^2 a$, $\beta^2 a$
and $\zeta^2 a$ respectively denote the probabilities of
fluctuations $u(d) \rightarrow s + K^{-(0)}$, $u(d,s) \rightarrow
u(d,s) + \eta$, and $u(d,s) \rightarrow u(d,s) + \eta^{'}$.

For the sake of simplification, the GB field can also be expressed
in terms of the quark contents of the GBs and is expressed as \bea
{\Phi'} = \left( \ba{ccc} \phi_{uu} u \bar u+ \phi_{ud} d \bar d
+\phi_{us} s \bar s& \varphi_{ud} u \bar d & \varphi_{us} u \bar s
\\ \varphi_{du} d \bar u & \phi_{du}u \bar u+ \phi_{dd} d \bar d
+\phi_{ds} s \bar s & \varphi_{ds} d \bar s
\\ \varphi_{su} s \bar u & \phi_{sd} s \bar d & \phi_{su} u \bar u
+ \phi_{sd} d \bar d +\phi_{ss} s \bar s \\ \ea \right)\,, \eea
where \bea \phi_{uu} &=& \phi_{dd}= \frac{1}{2} +\frac{\beta}{6} +
\frac{\zeta}{3}\,,~~~~~ \phi_{ss} =\frac{2\beta}{3} +
\frac{\zeta}{3}\,,~~~~~ \phi_{us} = \phi_{ds}=
\phi_{su}=\phi_{sd}= -\frac{\beta}{3} + \frac{\zeta}{3}\,,
\nonumber \\ \phi_{du} &=& \phi_{ud}= -\frac{1}{2}
+\frac{\beta}{6} + \frac{\zeta}{3}\,,~~~~~ \varphi_{ud} =
\varphi_{du} = 1\,, ~~~~\varphi_{us} = \varphi_{ds}=\varphi_{su}=
\varphi_{sd}= \alpha \,.  \eea The quark sea content of the baryon
can be calculated in $\chi$CQM by substituting for every
constituent quark $q \to \sum P_q q + |\psi(q)|^2$, where $\sum
P_q$ is the transition probability of the emission of a GB from
any of the $q$ quark and $|\psi(q)|^2$  is the transition
probability of the $q$  quark. The flavor structure for the baryon
of the type $B(xxy)$ is expressed as $2 P_x x + P_y y + 2
|\psi(x)|^2 + |\psi(y)|^2$ and for the  type $B(xyz)$ it is
expressed as $ P_x x + P_y y + P_z z+ |\psi(x)|^2 + |\psi(y)|^2+
|\psi(z)|^2$, where $x,y,z=u,d,s$. The sea quark distribution
function for the octet baryons $p$, $\Sigma^+$, $\Sigma^0$, and
$\Xi^0$ have been presented in Table \ref{antiquark}.

\begin{table}[h]
\begin{tabular}{|c|c|c|c|}
\hline Baryon & $\bar u$ & $\bar d$ & $\bar s$ \\\hline $p(uud)$ &
$a(2\phi_{uu}^2 + \phi_{du}^2 + \varphi_{du}^2)$ & $a(2\phi_{ud}^2
+2\varphi_{ud}^2 + \phi_{dd}^2) $ & $a(2\phi_{us}^2 +
2\varphi_{us}^2 + \phi_{ds}^2 + \varphi_{us}^2)$ \\

$\Sigma^{+}(uus)$ & $a(2\phi_{uu}^2 + \phi_{su}^2 +
\varphi_{su}^2)$ & $a(2\phi_{ud}^2 +2\varphi_{ud}^2 +\phi_{sd}^2
+\varphi_{sd}^2)$ & $a(2\phi_{us}^2+ 2\varphi_{us}^2+
\phi_{ss}^2)$ \\

$\Sigma^{0}(uds)$ & $a(\phi_{uu}^2 +\phi_{du}^2 + \phi_{su}^2 +
\varphi_{du}^2 + \varphi_{su}^2)$ & $a(\phi_{ud}^2 + \phi_{dd}^2 +
\phi_{sd}^2 + \varphi_{ud}^2 +\varphi_{sd}^2)$ & $a(\phi_{us}^2 +
\phi_{ds}^2 + \phi_{ss}^2 + \varphi_{us}^2+ \varphi_{ds}^2)$ \\
$\Xi^{0}(uss)$ &$ a(\phi_{uu}^2 +2 \phi_{su}^2 +2 \varphi_{su}^2)$
& $a(\phi_{ud}^2 +\varphi_{ud}^2+ 2\phi_{sd}^2 +2 \varphi_{sd}^2)
$ & $a(\phi_{us}^2 +\varphi_{us}^2+2\phi_{ss}^2)$  \\ \hline
\end{tabular}
\caption{The sea quark distribution functions for the octet
baryons. The expressions for other octet baryons can be
obtained through isospin symmetry.}\label{antiquark}
\end{table}

There are no simple or straightforward rules which could allow
incorporation of $x-$dependence in $\chi$CQM. To this end, instead
of using an {\it ab initio} approach, we have phenomenologically
incorporated the $x-$dependence getting clues from Eichten {\it et
al.} \cite{eichten}, Isgur \cite{Isgur} {\it et al.} and Le
Yaouanc {\it et al.} \cite{yaouanc}. The $x-$dependent sea quark
distribution functions can be now expressed as $\bar u^B(x)=\bar
u^B{(1-x)}^{10}$, $\bar d^B(x)=d^B{(1-x)}^{7}$, $\bar
s^B(x)=s^B{(1-x)}^{8}$ which together with the valence quark
distribution functions give the flavor structure of the baryon as
\be q^B(x)=q^B_{{\rm val}}(x)+\bar q^B(x)\,, \label{totalquark}\ee
where $q=u,d,s$. Using the sea quark distribution functions from
Table \ref{antiquark}, the quark sea asymmetries $\bar u(x) - \bar
d(x)$ and $\bar d(x)/\bar u(x)$ can also be calculated at
different $x$ values. We have already discussed the inclusion of
$x$-dependence in detail and compared our results with the
experimental data for the case of nucleon in Ref. \cite{xdep}. In
the present communication however, we have extended our
calculations to the case of other octet baryons for which
experimental data is not yet available.

The $x-$dependence of the structure functions $F_1$ and $F_2$ can
be calculated from \bea F_2^B(x) &=& x \sum_{u,d,s} e^2_{q}[q^B(x)
+ \bar q^B(x)] \,,\\ F_1^B(x) &=& \frac{1}{2x}F_2^B(x) \,, \eea
where $e_q$ is the charge of the quark $q$ ($e_u=\frac{2}{3}$ and
$e_d= e_s=-\frac{1}{3}$). In terms of the quark distribution
functions, the structure function $F_2$ for any baryon can be
expressed as \be F^B_2(x)=\frac{4}{9} x(u^B(x)+ \bar u^B(x))
+\frac{1}{9} x(d^B(x)+ \bar d^B(x)+ s^B(x)+\bar s^B(x))\,. \ee
Several important quantities can be obtained from the structure
functions of different isospin multiplets. For example, for the
case of proton and neutron we have \bea \frac{F^p_2(x)
-F^n_2(x)}{x} &=& \frac{4}{9}\left(u_{\rm val}^p(x) - u_{\rm
val}^n(x) + 2\bar u^p(x) - 2\bar u^n(x) \right) +\frac{1}{9}
\left( d_{\rm val}^p(x) + s_{\rm val}^p(x) \right. \nonumber \\
&-& \left. d_{\rm val}^n(x)- s_{\rm val}^n(x) + 2\bar d^p(x)+ 2
\bar s^p(x) - 2 \bar d^n(x)- 2 \bar s^n(x) \right). \label{ig}
\eea

The Gottfried integral $I^{p n}_G$ \cite{gsr} can be expressed in
terms of the sea quarks as follows \be I^{p n}_G = \int_0^1
{\frac{F^p_2(x) -F^n_2(x)}{x}} dx = \frac{1}{3} + \frac{2}{3}
\int_0^1 \left[ \bar u^p(x)- \bar d^p(x) \right] dx\,,
\label{ipn}\ee where we have used the following normalization
conditions \[ \int_0^1 u_{\rm val}^p(x)dx = 2 \,,~~ \int_0^1 d_{\rm
val}^p(x)dx=1 \,, \int_0^1 s_{\rm val}^p(x)dx=0 \,, \]
\[\int_0^1 u_{\rm val}^n(x)dx=1 \,, \int_0^1 d_{\rm val}^n(x)dx=2
\,, \int_0^1 s_{\rm val}^n(x)dx=0 \,, \] \be \int_0^1 \bar
d^n(x)dx= \int_0^1 \bar u^p(x) dx \,,~~  \int_0^1 \bar u^n(x)dx =
\int_0^1 \bar d^p(x) dx \,,~~ \int_0^1 \bar s^n(x)dx= \int_0^1
\bar s^p(x) dx \,. \ee
Similarly, for the case of other octet baryons the following normalization conditions  \bea \int_0^1 u_{\rm
val}^{\Sigma^+}(x)dx = 2 \,,~~~ \int_0^1 d_{\rm
val}^{\Sigma^+}(x)dx = 0 \,,~~~ \int_0^1 s_{\rm val}^{\Sigma^+} dx
= 1 \,, \nonumber \\ \int_0^1 u_{\rm val}^{\Sigma^0}(x)dx = 1
\,,~~~ \int_0^1 d_{\rm val}^{\Sigma^0}(x) dx=1 \,,~~~ \int_0^1
s_{\rm val}^{\Sigma^0}(x) dx = 1 \,, \nonumber\\ \int_0^1 u_{\rm
val}^{\Xi^{0}}(x) dx = 1 \,,~~~ \int_0^1 d_{\rm val}^{\Xi^{0}}(x)
dx = 0 \,,~~~ \int_0^1 s_{\rm val}^{\Xi^{0}}(x) dx = 2 \,,
\label{norm-s} \eea lead to the other Gottfried integrals in terms
of the sea quarks \bea I^{\Sigma^+ \Sigma^0}_G \equiv \int_0^1
\frac{F^{\Sigma^+}_2(x) -F^{\Sigma^0}_2(x)}{x} dx &=& \frac{1}{3}
+ \frac{2}{9} \int_0^1 \left[ 4 \bar u^{\Sigma^+}(x)+ \bar
d^{\Sigma^+}(x) - 4 \bar u^{\Sigma^0}(x) -\bar d^{\Sigma^0}(x)
\right] dx \,,\nonumber\\ I^{\Sigma^0 \Sigma^-}_G \equiv \int_0^1
\frac{F^{\Sigma^0}_2(x) -F^{\Sigma^-}_2(x)}{x} dx &=& \frac{1}{3}
+ \frac{2}{9} \int_0^1 \left[ 4 \bar u^{\Sigma^0}(x)+ \bar
d^{\Sigma^0}(x) - 4 \bar d^{\Sigma^+}(x) - \bar u^{\Sigma^+}(x)
\right] dx\,,\nonumber\\  I^{\Xi^{0} \Xi^{-}}_G \equiv \int_0^1{
\frac{F^{\Xi^0}_2(x) -F^{\Xi^-}_2(x)}{x}} dx &=& \frac{1}{3} +
\frac{2}{3} \int_0^1 \left[\bar u^{\Xi^0}(x) - \bar d^{\Xi^0}(x)
\right] dx\,. \label{ioctet}\eea It is clear from Eqs. (\ref{ipn})
and (\ref{ioctet}), the flavor symmetric sea leads to the
Gottfried sum rule $I_G=\frac{1}{3}$ with $\bar u^B$=$\bar d^B$.

\begin{figure}
\includegraphics[width=8cm,height=7cm,clip]{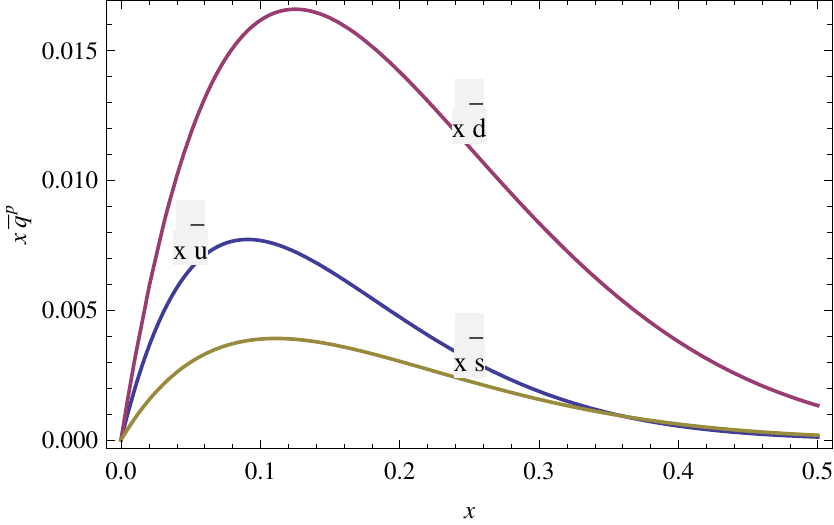}
\hspace{0.2cm}
\includegraphics[width=8cm,height=7cm]{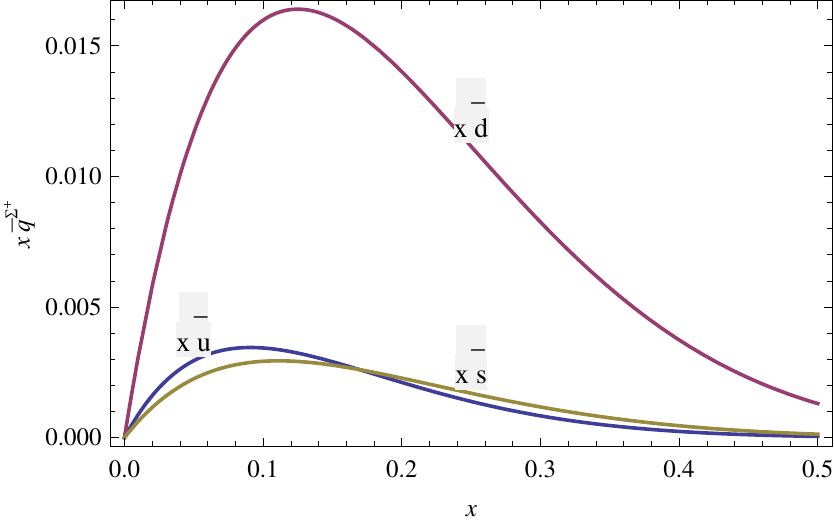}
\hspace{0.2cm}
\includegraphics[width=8cm,height=7cm]{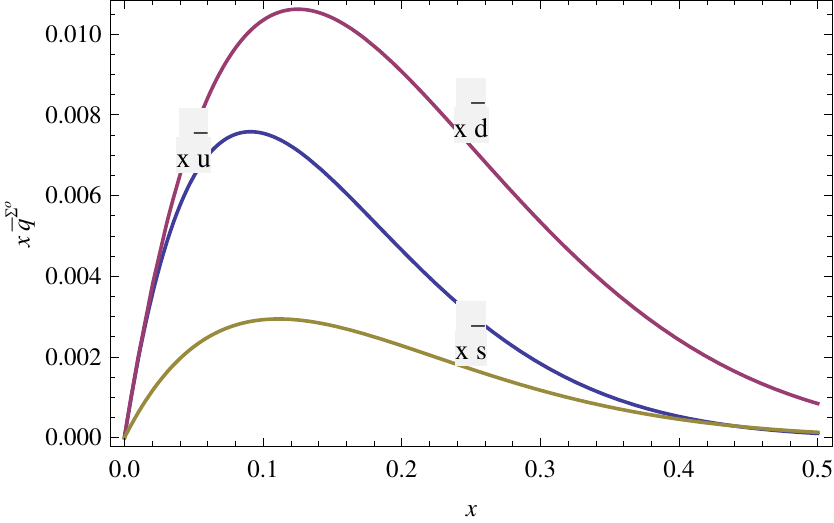}
\hspace{0.2cm}
\includegraphics[width=8cm,height=7cm]{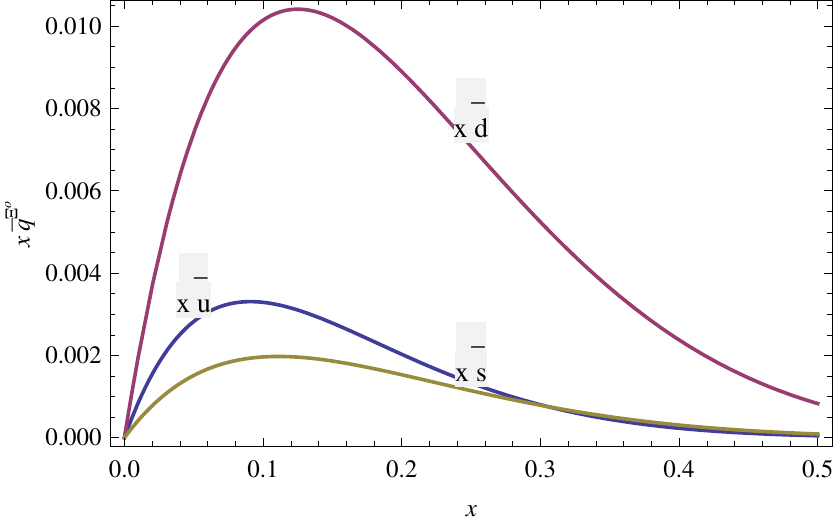}
\caption{Sea quark distribution functions as a function of Bjorken
scaling variable $x$ for the $p$, $\Sigma^+$, $\Sigma^0$, and $\Xi^0$, respectively.} \label{udsoctet}
\end{figure}

After having detailed the contribution of the quark sea and the
various asymmetries in the octet baryons of different quark
structure, we now discuss the variation of these quantities with
the Bjorken  variable $x$. For the numerical calculation of the
sea quark distribution functions of the octet baryons, we have
used the same set of input parameters as detailed in our earlier
calculations \cite{pdg,hd,xdep,ns}. In Fig. \ref{udsoctet}, we
have shown the variation of the sea quark distributions $x \bar
u(x)$, $x \bar d(x)$ and $x \bar s(x)$ with the Bjorken scaling
variable $x$ for $p(uud)$, $\Sigma^+(uus)$, $\Sigma^0(uds)$ and
$\Xi^0(uss)$. From a cursory look at the plots, one can easily
find out that

\[\bar d^{p}(x)>\bar u^{p}(x)>\bar s^{p}(x),\]
\[\bar d^{\Sigma^+}(x)>\bar u^{\Sigma^+}(x)\approx\bar s^{\Sigma^+}(x),\]
\[\bar d^{\Sigma^0}(x)>\bar u^{\Sigma^0}(x)>\bar s^{\Sigma^0}(x),\]
\[\bar d^{\Xi^0}(x)>\bar u^{\Xi^0}(x)>\bar s^{\Xi^0}(x),\]
showing a clear quark sea asymmetry as observed in the DIS
experiments \cite{nmc,e866,hermes}. These distributions clearly
indicate that our results pertaining to the quark sea asymmetry
seem to be well in line with the expected results.

A careful study of the plots brings out several interesting
points. As already mentioned in the introduction, the sea quarks
do not contribute at higher values of $x$, therefore in Fig.
\ref{udsoctet}, we have taken the region $x=0-0.5$. Beyond this
$x$ region the contribution  of the sea quarks is negligible and
the contributions should be completely dominated by the valence
quarks. The difference between the various sea distributions is
observed to be maximum at $x\approx0.1$. As the value of $x$
increases, the difference between the sea contributions decreases
in all the cases which is in line with the observations of other
models \cite{cqsm,stat,bag,alwall}.

\begin{figure}[t]
\includegraphics[width=8cm,height=7cm]{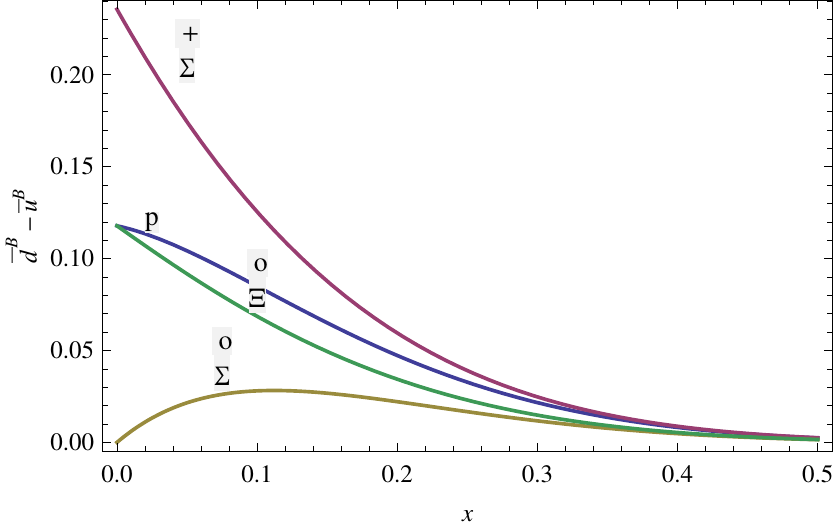}
\hspace{0.2cm}
\includegraphics[width=8cm,height=7cm]{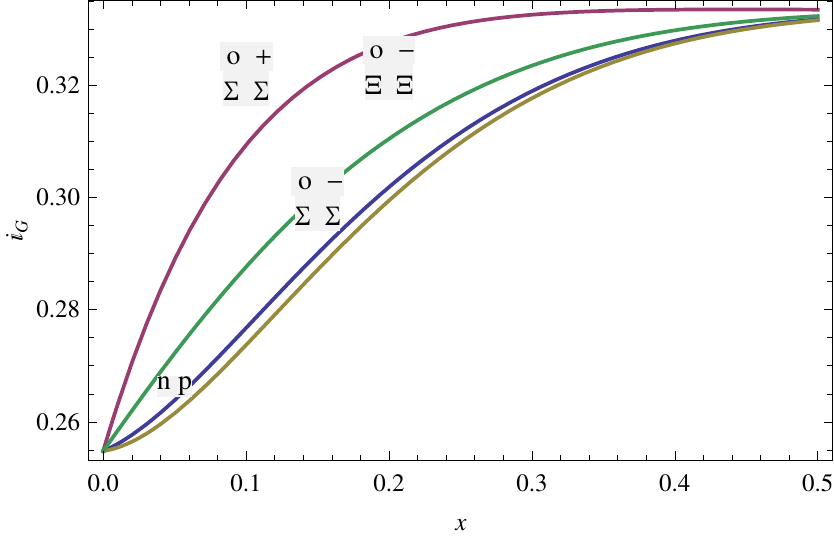}
\caption{Sea quark asymmetry and Gottfried integrals as a function of Bjorken scaling variable $x$ for the
octet baryons.} \label{bdu}
\end{figure}

The general aspects of the variation of the magnitudes of the sea
quark distribution functions $\bar u(x)$, $\bar d(x)$ and $\bar
s(x)$ for the octet baryons are able to explain some of the well
known experimentally measurable quantities, for example, $\bar
d^{B}(x)-\bar u^{B}(x)$, $\bar d^{B}(x)/\bar u^{B}(x)$ and the
Gottfried integral. These quantities not only provide important
constraint on a model that attempts to describe the origin of the
quark sea but also provide a direct determination of the presence
of significant amount of quark sea in the low $x$ region. In Fig.
\ref{bdu}, the $\chi$CQM results for the $\bar
d^{B}(x)-\bar u^{B}(x)$ and the Gottfried integrals have been
plotted at different $x$ values. It is clear from the plots that
when $x$ is small $\bar d^{B}(x)-\bar u^{B}(x)$ asymmetries are
large implying the dominance of sea quarks in the low $x$ region.
In fact, the sea quarks dominate only in the region where $x$ is
smaller than 0.3. At the values $x>0.3$, $\bar d-\bar u$ tends to
0 implying that there are no sea quarks in this region. The
contribution of the quark sea in the case of $\Sigma^0$ is
particularly interesting because of its flavor structure which has
equal numbers of $u$, $d$ and $s$ quarks in its valence structure.
Unlike the other octet baryons, where the $\bar d(x)-\bar u(x)$
asymmetry decreases continuously with the $x$ values, the
asymmetry in this case first increases and then for values of
$x>0.1$ it decreases. However, it is interesting to observe the
the asymmetry peak in this case which matches with our other
predictions where the contribution of the quark sea is maximum at
$x\approx0.1$

A measurement of the Gottfried integral \cite{nmc,e866} for the
case of nucleon has shown a clear violation of Gottfried sum rule
from $\frac{1}{3}$  which can find its explanation in a global
quark sea asymmetry $ \int_0^1 (\bar d(x) -\bar u(x))dx$.
Similarly, for the case of $\Sigma^{+}$, $\Sigma^{0}$, and
$\Xi^0$, the Gottfried sum rules should read $I^{\Sigma^+
\Sigma^0}_G=\frac{1}{3}$, $I^{\Sigma^0 \Sigma^-}_G=\frac{1}{3}$
and $I^{\Xi^0 \Xi^-}_G=\frac{1}{3}$ if the quark sea was
symmetric. However, due to the $\bar d(x)-\bar u(x)$ asymmetry in
the case of octet baryons, a lower value of the Gottfried
integrals is obtained. We have plotted the results in Fig.
\ref{bdu}. In the case of nucleon the results are in good
agreement with the experimental data \cite{e866} as already
presented in \cite{xdep}. The quality of numerical agreement in
the other cases can be assessed only after the data gets refined.
Further, this phenomenological analysis strongly suggests an
important role for the quark sea at low value of $x$. New
experiments aimed at measuring the flavor content of the other
octet baryons are needed for profound understanding of the
nonperturbative properties of QCD.

To summarize, in order to investigate the effects of ``quark
sea'', we have calculated the sea quark distribution functions for
the octet baryons  in the chiral constituent quark model
($\chi$CQM). The Bjorken scaling variable $x$ has been
incorporated phenomenologically to enlarge the scope of model and
to understand the range of $x$ where quark sea effects are
important. Implications of the quark sea have also been studied to
estimate the quark sea asymmetries like $\bar d(x)-\bar u(x)$,
$\bar d(x)/\bar u(x)$ and Gottfried integral. The results justify
our conclusion regarding  the importance of quark sea  at small
values of $x$.

In conclusion, the results obtained for the quark distribution
functions reinforce our conclusion that $\chi$CQM is able to
generate qualitatively as well as quantitatively the requisite
amount of quark sea. This can perhaps be substantiated by a
measurement of the quark distribution functions of the other octet
baryons.

\vskip .2cm {\bf ACKNOWLEDGMENTS}\\ H.D. would like to thank
Department of Science and Technology, Government of India, for
financial support.

\end{document}